\newif\ifmnras
\mnrasfalse
\ifmnras
	\documentclass[useAMS,usenatbib]{mn2e}
\else
	\documentclass[apj,numberedappendix]{emulateapj}
\fi
\usepackage{times}
\usepackage{graphics,epsf}
\usepackage{amsmath}                
\usepackage{amsfonts}               
\usepackage{amssymb}                
\usepackage{epsfig}                 
\usepackage{graphicx}
\usepackage{url}
\usepackage{booktabs}
\usepackage{hyperref}
\usepackage{color}
\def \kms{\rm{km}$\rm{s}^{-1}$}

\def \Jcgs{$\rm{g}~\rm{cm}^{2}~\rm{s}^{-1}$}
\def \jcgs{$\rm{cm}^{2}~\rm{s}^{-1}$}

\def \h{~\rm{hr}}
\def \s{~\rm{s}}

\def \km{~\rm{km}}
\def \kms{~\rm{km}~{\rm s}^{-1}}

\def \K{~\rm{K}}

\def \erg{~\rm{erg}}
\def \foe{~\rm{foe}}

\def \yr{~\rm{yr}}

\def \days{~\rm{day}}

\def \foe{~\rm{foe}}

\def \zams{\mathrm{ZAMS}}

\ifmnras
	\def \aap{A\&A}
	\def \apj{ApJ}
	\def \apjl{ApJ}
	\def \apjs{ApJS}
	\def \apss{AP\&SS}
	\def \araa{ARA\&A}
	\def \nat{Nature}
	\def \mnras{MNRAS}
\fi

\def \newa{NewA}
\def \pasa{PASA}
\def \raa{RAA}

\ifmnras
	\title[Explaining energetic supernovae with an inefficient JFM]{Explaining the most energetic supernovae with an inefficient jet-feedback mechanism}

	\author[A. Gilkis, N. Soker \& O. Papish]{Avishai Gilkis, Noam Soker and Oded Papish\\
	Department of Physics, Technion -- Israel, Institute of Technology, Haifa 32000, Israel;\\
	agilkis@tx.technion.ac.il;
	soker@physics.technion.ac.il;
	papish@campus.technion.ac.il}
\fi

\begin{document}

\ifmnras
	\pagerange{\pageref{firstpage}--\pageref{lastpage}} \pubyear{2015}

	\maketitle
\else
	\title{Explaining the most energetic supernovae with an inefficient jet-feedback mechanism}

	\author{Avishai Gilkis}
	\author{Noam Soker}
	\author{Oded Papish}
	\affil{Department of Physics, Technion -- Israel
	Institute of Technology, Haifa 32000, Israel;
	agilkis@tx.technion.ac.il; soker@physics.technion.ac.il; papish@campus.technion.ac.il}
\fi


\label{firstpage}

\begin{abstract}
We suggest that the energetic radiation from
core-collapse super-energetic supernovae (SESNe)
is due to a long lasting accretion process onto the newly born neutron star (NS),
resulting from an inefficient operation of the jet-feedback mechanism.
The jets that are launched by the accreting NS
or black hole (BH) maintain their axis due to a rapidly rotating pre-collapse core,
and do not manage to eject core material from near the equatorial plane.
The jets are able to eject material from the core along the polar directions,
and reduce the gravity near the equatorial plane.
The equatorial gas expands,
and part of it falls back over a timescale of minutes to days
to prolong the jets-launching episode.
According to the model for SESNe proposed in the present paper,
the principal parameter that distinguishes between the different cases of CCSN explosions,
such as between normal CCSNe and SESNe,
is the efficiency of the jet-feedback mechanism.
This efficiency in turn depends on the pre-collapse core mass,
envelope mass, core convection, and most of all on the angular momentum profile in the core.
One prediction of the inefficient jet-feedback mechanism for
SESNe is the formation of a slow equatorial outflow in the explosion.
Typical velocity and mass of this outflow are estimated to be
$v_\mathrm{eq} \approx 1000 \kms$ and
$M_\mathrm{eq} \ga 1 M_\odot$, respectively,
though quantitative values will have to be checked in
future hydrodynamic simulations.
\end{abstract}

\section{INTRODUCTION}
\label{sec:introduction}

In core-collapse supernova (CCSN) explosions of massive stars,
more than $10^{53} \erg$ of gravitational energy is released by the
collapse of the stellar core forming a neutron star (NS) or a black hole (BH).
There is as yet no consensus on the mechanism
that channels a small fraction of this energy to the kinetic
energy of the stellar ejecta, composed of the rest of the core and
the stellar envelope.
Most well-studied is the ``delayed neutrino-heating mechanism''
(\citealt{Wilson1985, Bethe1985};
see review by \citealt{Janka2012} and references therein),
while other approaches are the jittering jets mechanism
\citep{Soker2010, PapishSoker2011, PapishSoker2012a, PapishSoker2012b, PapishSoker2014a, PapishSoker2014b, GilkisSoker2014, GilkisSoker2015, Papishetal2015a}
and the collapse-induced thermonuclear explosion
(CITE; \citealt{Burbidgeetal1957, KushnirKatz2014, Kushnir2015a, Kushnir2015b}; see analysis in appendix \ref{app:cite}).

Although the increasingly sophisticated multidimensional simulations of the delayed neutrino mechanism
(e.g.,
\citealt{Burrows1985,Burrows1995,Fryer2002,Buras2003,
Ott2008,Marek2009,Nordhaus2010,Brandt2011,Hanke2012,Kuroda2012,
Hanke2012,Mueller2012,Bruennetal2013,Bruennetal2014,
CouchOtt2013,CouchOtt2015,
MuellerJanka2014,MuellerJanka2015,
Mezzacappaetal2014,Mezzacappaetal2015,Abdikamalovetal2015})
are much more advanced than those of the jittering jets mechanism \citep{PapishSoker2012a,PapishSoker2012b,PapishSoker2014a,PapishSoker2014b},
they have on the other hand failed to explode CCSNe at the desired energies.
Still,
current state-of-the art neutrino-driven simulations
are not converged and there is no final word yet (see \citealt{Janka2016} for a review).
The delayed neutrino mechanism faces first the problem of reviving the stalled shock of the inflowing gas
(e.g., \citealt{Fernandez2015, Gabayetal2015} for recent papers),
and then the challenge of achieving an explosion kinetic energy of
$E_\mathrm{kin} \approx 1-10 \foe$, where $\foe \equiv 10^{51} \erg$.
As explained by \cite{Papishetal2015b},
even if the stalled shock is revived,
there is a generic problem in the delayed neutrino mechanism that limits the final kinetic energy to
$E_\mathrm{kin} \la 0.2-0.5 \foe$,
and only when the energy is scaled in advance do the simulations get to energies of $\ga 0.5 \times 10^{51} \erg$
(e.g., \citealt{Sukhboldetal2016}).
A more efficient neutrino-driven explosion might take place when it is enhanced by convection
(`convection-engine'; \citealt{Fryer2006}).
Still the explosion energy will be limited to about $2 \foe$ \citep{Fryeretal2012},
or at most $3 \foe$ \citep{SukhboldWoosley2016}.
Some other problems that the delayed neutrino mechanism faces are listed by \cite{Kushnir2015b}.

The jet-feedback mechanism (JFM),
including the jittering jets model and cases with a constant jets axis,
assumes that in all CCSNe with energies above $0.5 \times 10^{51} \erg$,
from ordinary to exotic, the dominating powering is by jets.
The main challenge of the jittering jets model
is  to supply a large enough specific angular momentum to the mass
accreted onto the NS or BH to form an accretion disk or an accretion
belt in all CCSNe \citep{GilkisSoker2014, GilkisSoker2015, GilkisSoker2016}.
The problem of forming an accretion disk is lessened when the core acquires a large amount of angular momentum,
e.g., by specific types of binary interaction
(e.g., \citealt{Izzardetal2004, Podsiadlowskietal2004, FryerHeger2005, Yoonetal2010}).
The second major challenge is the formation mechanism of jets.
Although this challenge is shared with jet-driven models in rare cases with pre-collapse rapidly-rotating cores
(e.g. \citealt{LeBlanc1970, Khokhlov1999, Lazzati2012}),
the jittering jets model may require jet formation to operate for more modest core rotation
(see, e.g., \citealt{Akiyama2003, SchreierSoker2016} for possible mechanisms).
Full magnetohydrodynamics simulations of collapsing rapidly-rotating strongly magnetized cores
(e.g., \citealt{Mostaetal2014})
still lack the required to resolve the magnetorotational instability \citep{Rembiasz2016},
and this is an issue of ongoing study.

Recent surveys have discovered many cases of super-energetic supernovae (SESNe),
and the question of their origin and their powering mechanism has become a hot unsolved topic
(e.g., \citealt{Quimbyetal2011, Quimbyetal2013, Moriyaetal2015,
Arcavietal2016, Perleyetal2016, Sorokinaetal2016}; see review by \citealt{GalYam2012}).
\cite{Sandersetal2012}, for example, discuss SN~2010ay, a broad-line Type Ic SN (Type Ic-BL SN).
They estimate the ejected mass to be $M_\mathrm{ej} \approx 4.7 M_\odot$,
and the kinetic energy to be $E_\mathrm{kin} \approx 1.1 \times 10^{52} \erg$.
Another extreme case is the hydrogen-poor SESN ASASSN-15lh reported by \cite{Dongetal2015},
where the total radiated energy is estimated to be $E_\mathrm{rad} \approx 7.5 \times 10^{51} \erg$.
\cite{Dongetal2015} argue that this most luminous SN discovered
challenges all theoretical scenarios for its origin,
although they do not rule out these scenarios.

In the delayed neutrino mechanism,
the explosion energy does not reach the observed values of the super-energetic cases,
even in scaled simulations such as those of \cite{Sukhboldetal2016}.
One alternative model for SESNe is based on quark-nova (e.g., \citealt{Ouyedetaletal2015, Ouyedetaletal2016}).
Another one is based on the formation of a magnetar
\citep{KasenBildsten2010, Woosley2010, Metzgeretal2015}.
\cite{Daietal2015}
propose that ASASSN-15lh could have been powered by
a newborn strongly-magnetized pulsar rotating with a nearly Keplerian period
(see also \citealt{Metzgeretal2015} and \citealt{Berstenetal2016}).
\cite{Wangetal2015} propose that the light curve of the SESN iPTF13ehe
has been powered by the three mechanisms of radioactive nickel, a magnetar,
and collision with circumstellar matter (CSM).
They estimate the total kinetic energy of this SESN according to their fit to be
$E_\mathrm{kin} \approx 3.5 \times 10^{52} \erg$.
The energy in their proposed magnetar cannot supply this kinetic energy.
However, \cite{Metzgeretal2015} give a new estimate of the energy
that can be supplied by magnetars of up to $\approx 10^{53} \erg$,
possibly explaining all SESNe.
\cite{SukhboldWoosley2016} also discuss the limits on radiated energy in SESNe,
giving an upper limit of $4 \times 10^{52} \erg$ for magnetar-powered Type I SNe.

\cite{BarkovKomissarov2011} raised the idea that a NS that is
spun-up in a common envelope evolution with a giant can lead to energetic SNe.
\cite{FryerWoosley1998} suggested that a merger of a BH with a helium core of a giant star can lead to a gamma-ray burst (GRB),
but they mention no jets.
\cite{Chevalier2012} proposed that a NS or a BH companion spiraling
inside the envelope can accrete mass at very high rates,
launch jets, and lead to a very energetic SN.
\cite{Papishetal2015c} further discuss this scenario and raise
the possibility that strong r-process nucleosynthesis,
where elements with high atomic weight of $A \ga 130$ are formed,
occurs inside the jets launched by the NS.

The magnetar formation discussed above requires strong magnetic fields as well as high specific angular momentum,
to the point that an accretion disk may be formed.
These are the two ingredients leading to very efficient launching of jets in many other astrophysical objects,
from young stellar objects to active galactic nuclei (e.g., \citealt{Livio2004}).
The possibility that magnetar formation is accompanied by jets should be considered \citep{Soker2016}.
Although the exact details of jet formation remain undetermined,
the above considerations, and the presence of jets in long gamma-ray bursts
(LGRBs; e.g. \citealt{ShavivDar1995,Sari1999,Granot2014,KumarZhang2015}),
lead us to consider jets as the main powering mechanism of \textit{all} energetic SNe,
i.e., a total kinetic energy of $E_\mathrm{kin} \ga 0.5 \foe$.

In the present study we apply the JFM to SESNe,
suggesting a CCSN mechanism spanning a large range of explosion energies
where jets are the primary driving power,
with a varying efficiency of envelope expulsion
from the efficient jittering jets regime to the inefficient steady jets limit.
In section \ref{sec:jittering} we present the jittering jets
mechanism, and argue that it can account for any SESN in terms of energetics.
In section \ref{sec:asymmetricexploison} we discuss the outcome
of explosions set up by weakly or non jittering jets,
hence a limited feedback process.
Our summary is brought in section \ref{sec:summary}.

\section{THE JITTERING JETS MECHANISM}
\label{sec:jittering}

In the jittering jets mechanism the newly born NS is assumed to launch jets with varying directions
\citep{PapishSoker2011}.
The jets are shocked within the core and the hot shocked gas inflates hot bubbles,
that might merge to one large bubble.
Because of the varying directions, the jets quite efficiently
eject the rest of the core and the entire envelope of the star.
When the jets manage to eject the core material, they shut themselves
out, in a negative feedback mechanism. The operation of the
jittering jets mechanism in an efficient negative feedback process
accounts for the explosion energy being several times the binding
energy of the core.

The directional variations of the jets arise from
the stochastic angular momentum of the accreted gas as depicted in Figure \ref{fig:schem1}.
Possible sources of the stochastic angular momentum of the accreted mass are
the convection in the pre-collapse envelope \citep{GilkisSoker2014, GilkisSoker2015, GilkisSoker2016},
and the spiral mode of the standing accretion shock instability (SASI; \citealt{Papishetal2015a}).
The SASI modes have been studied extensively in the past
(e.g., \citealt{BlondinMezzacappa2003, BlondinMezzacappa2007, Fernandez2010, Burrows1995, Janka1996, Buras2006a, Buras2006b, Ott2008, Marek2009}),
but these are the spiral modes of the SASI studied in recent years
(e.g., \citealt{Fernandez2010, Hanke2013, Fernandez2015, Kazeronietal2015})
that seem most promising as a source of stochastic angular momentum
for the jittering jets model \citep{Papishetal2015a}.
For jets to form from these instabilities in numerical simulations,
pre-collapse conditions which include the turbulent flow of convective regions
(e.g., \citealt{Chatzopoulos2014,Chatzopoulos2016,Couchetal2015})
are required for the case of convection as a source of angular momentum,
and realistic magnetic fields and their amplification during collapse for both.
The combined problems of uncertain initial conditions and insufficient resolution
limit the feasibility of fully simulating the jittering jet scenario.
\ifmnras
\begin{figure}
\else
\begin{figure}[ht!]
\fi
   \centering
   \ifmnras
    \includegraphics*[scale=0.41]{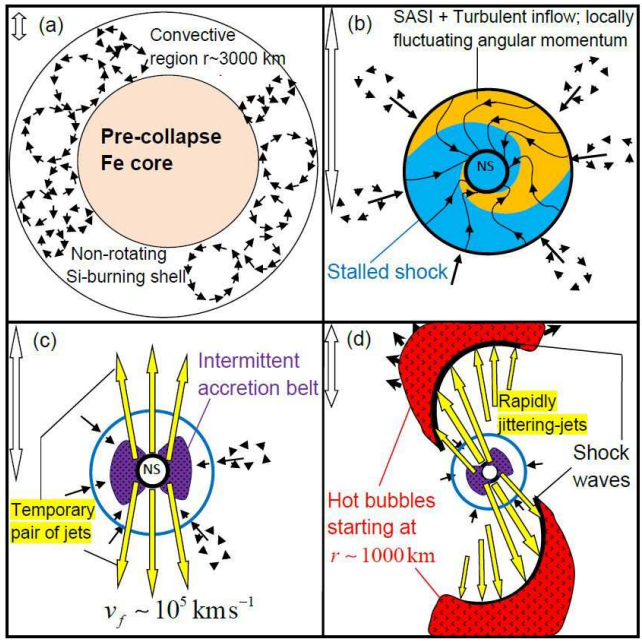} \\
   \else
    \includegraphics*[scale=0.42]{SLSNjetsFigure1l.eps} \\
    \fi
\caption{A schematic presentation of the jittering jets
mechanism in a non-rotating (or slowly rotating) core, spanning an
evolution time of several seconds (taken from
\citealt{Papishetal2015a}). The two-sided arrow on the upper left
of each panel corresponds to a length of approximately $500 \km$.
(a) The convective vortices in the silicon burning shell of the
pre-collapse core are a source of stochastic angular momentum. (b)
After collapse and the formation of a NS the rest of the
in-falling gas passes through the stalled shock, and the spiral
modes of the standing accretion shock instability (SASI) add to
the stochastic angular momentum of the accreted gas. (c) As a
result of the stochastic nature of the angular momentum of the
accreted gas, both in magnitude and in direction, for short
periods of times, tens of milliseconds, an intermittent accretion
belt is formed around the newly born NS. If the disk exists for a
long enough time, several dynamical times, or $> 0.01 \s$, it can
spread in the radial direction to form an accretion disk. The
disks (or belts) are assumed to launch two opposite jets with
initial velocities of $v_f \approx 10^5 \km \s^{-1}$ (about the
escape velocity from the newly formed NS). (d) The jittering jets
that are launched in varying directions inflate hot bubbles (see
\citealt{PapishSoker2011}). These bubbles expand and explode the star
in the jittering jets model \citep{PapishSoker2014a,
PapishSoker2014b}.}
      \label{fig:schem1}
\end{figure}

An interesting case is the $^{44}Ti$ distribution in the SN remnant Cassiopeia~A.
\cite{Grefenstetteetal2014} map this distribution and
argue against fast-rotating progenitors as well as jet-like explosions,
instead suggesting that the $^{44}Ti$ non-uniformity is the result of a less symmetric explosion mechanism,
termed multi-modal explosion, such as expected from instabilities (e.g., the SASI; \citealt{Fernandez2015}).
The jittering jets mechanism also has the property of multi-modal explosion.
If several pairs of opposite jets are launched in different directions and explode the star,
then the explosion has no symmetry axis or symmetry plane.
After the core explosion by the earlier jets,
a last episode of bipolar jets may propagate freely in the inner region and leave an imprint on the outer ejecta.
We speculate that the ``jet'' structure observed in 
Cassiopeia~A and its counter protrusion are the result of a last jets-launching episode in the context of the jittering jets model,
but further study will have to address this case in detail.

The explosion energy comes from the jets launched by the accretion disk around the newly formed NS (or BH).
Typically, about $10 \%$ of the accreted mass after the NS was born, $M_\mathrm{acc}$,
is assumed to be launched at the escape velocity of $\approx 0.3-0.5 c$.
The rest of the accretion energy is lost in neutrino cooling.
The jet-launching starts shortly after the bounce of the shock,
or when the central mass is approximately $1.2 M_\odot$.
The energy that is channeled from the accreted gas to the jets is therefore
$E_\mathrm{tot} \simeq 0.5 M_\mathrm{jet} (0.4 c)^2$,
where the mass carried by the jets is $ M_\mathrm{jet} \simeq 0.1 M_\mathrm{acc}$.
The accreted mass in units of solar masses is $M_\mathrm{acc}=M_\mathrm{rem} -1.2$,
where $M_\mathrm{rem}$ is the final mass in solar masses of the remnant (a NS or a BH).
According to these considerations,
the jittering jets model can account for a total explosion energy,
radiation plus kinetic,
of up to
\begin{equation}
 E_\mathrm{tot} = E_\mathrm{rad} + E_\mathrm{kin} \simeq 10
 \left( M_\mathrm{rem} - 1.2 \right) \foe.
 \label{eq:Etot}
\end{equation}
This can well account for all CCSNe,
even for very energetic SESNe when the remnant mass is large,
i.e., a BH is formed.

The jittering jets model strongly disfavors failed CCSNe.
To the contrary, if the jets launched by the inner $\approx 2-2.5
M_\odot$ of the core as it collapses to form a NS do not manage to
explode the star, then a BH is formed and a more violent supernova
explosion takes place, rather than a very low energy supernova
\citep{GilkisSoker2014}.
There are claims for missing CCSNe from progenitors of initial mass $\ga 20 M_\odot$, e.g.,
\cite{Smartt2015}, and \cite{Reynoldsetal2015} find a candidate for a failed CCSN.
If holds true, then the jittering jets model will have to be reevaluated.
On the other hand, there are claims in the last-year literature for very massive CCSN progenitors.
\cite{Strolgeretal2015} claim that CCSN progenitors have initial masses in the range $8-50 M_\odot$,
and there is no upper limit of $\approx 20 M_\odot$ on CCSN progenitors.
\cite{Nicholletal2015} find that some H-poor super-luminous CCSNe ejected mass of up to $30 M_\odot$.
Since they are H-poor, hence some envelope has been removed,
the initial mass of the progenitor was larger even.

The binding energy of the exploding part of the star in many CCSNe,
in particular those with an initial mass of above about $11 M_\odot$,
is $E_\mathrm{bind} \simeq \mathrm{few} \times 0.1 \foe$
(see Fig. \ref{fig:models1} in Appendix \ref{app:mesa}).
In the jittering jets model the negative feedback mechanism is efficient,
such that the total energy carried by the jets that removes the core and shut off the jets is
$E_\mathrm{jets} \approx (3-5) \times E_\mathrm{bind} \approx 1\foe$.
This may explain the peak in CCSN energy around $\approx 1\foe$.

In the present study we consider rare CCSNe where the feedback mechanism is inefficient.
The remnant mass and explosion energy depend on the efficiency
of the jets in removing core material.
High efficiency will result in a regular SN explosion energy
and a NS remnant.
Lower efficiency will lead to BH formation,
with typical masses of $M_\mathrm{BH}
\approx 3-10 M_\odot$ resulting in explosion energies in the range of
$E_\mathrm{tot} (\mathrm{BH}) \approx 30-100 \foe$.
This energy is similar to what magnetars can supply \citep{Metzgeretal2015}.

\section{INEFFICIENT JET-FEEDBACK MECHANISM}
\label{sec:asymmetricexploison}

\subsection{The effect of rapid rotation}
\label{subsec:inefficient}

We here discuss the case of a pre-collapse rapidly spinning core.
The effects of high core rotation have been studied in the past
(e.g., \citealt{Burrowsetal2007,Mostaetal2015}),
but we focus here on the aspect of inefficient feedback of such cases.

The gas collapsing onto the newly born NS has a large specific
angular momentum with a well defined axis.
Pre-collapse convection and post-collapse instabilities change the
angular momentum axis by small angles only.
Fast rotation can further decreases these fluctuations because the convection above the newly born NS is less vigorous
\citep{FryerHeger2000},
hence the specific angular momentum of the accreted mass is less stochastic.
The accretion disk is expected to last for the entire duration of the accretion process,
and to maintain the same plane.
The jets that are launched by the accretion disk have a well-defined and invariable axis.
The jet-inflated bubbles remove two opposite regions from the two
sides of the equatorial plane as depicted schematically in
Figure \ref{fig:schem2}.
We address the evolution of the core material near the equatorial plane,
which is not removed directly by the jets,
up to somewhere in the helium shell.
Before collapse and before mass removal this gas is in hydrostatic equilibrium.
This does not hold anymore after two dense regions have been removed
(lower right panel in Fig. \ref{fig:schem2}).
This has the interesting effect of setting an outward motion near the equatorial plane,
on which we elaborate henceforward.
\ifmnras
\begin{figure}
\else
\begin{figure}[ht!]
\fi
   \centering
   \ifmnras
   \includegraphics*[scale=0.41]{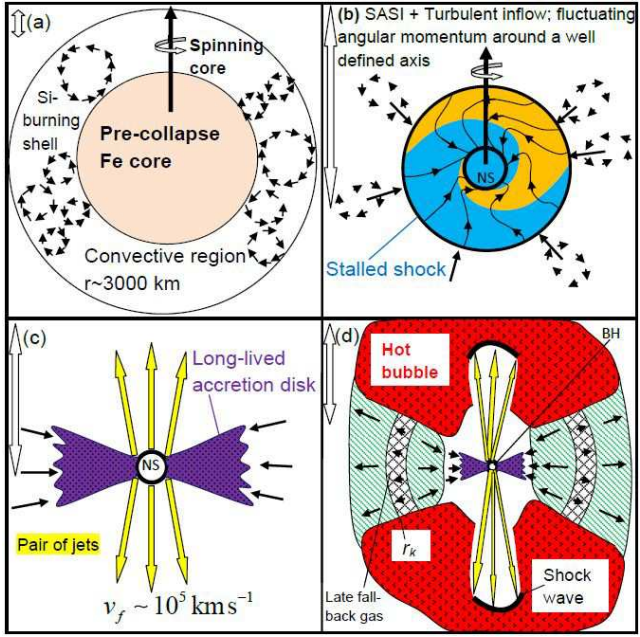} \\
   \else
   \includegraphics*[scale=0.42]{SLSNjetsFigure2l.eps} \\
   \fi
\caption{A schematic presentation of the inefficient feedback
mechanism operating in a weakly, or non, jittering jets case. (a)
The main source of angular momentum is a rapidly spinning
pre-collapse core. (b)+(c) The high specific angular momentum of
the accreted gas forms a long-lived accretion disk around the
newly born NS or BH. Convection and instabilities lead to a small
jittering of the disk and hence the launched jets. However, the
jets preserve their general axis. The two-sided arrow on the upper
left of the first three panels corresponds to a length of
approximately $500 \km$. (d) The well collimated and constant-axis
jets inflate two bubbles (red) and remove two large regions from
the dense core above and below the equatorial plane. The core gas
that is left near the equatorial plane (hatched green region)
feels a reduced gravitational attraction after the removal of
that gas. Its outer parts start to expand outward due to a pressure
gradient. Inner regions continue to flow inward due to a rarefaction
wave that propagates out. A particular region of the core (two
cross-hatched arcs in the figure, a ring in 3D) at a radius marked $\approx r_k$, is accelerated
outward, but not to the escape speed. In a time much longer than
the dynamical time, this `sluggish' gas can fall back and power
jets minutes to days after explosion. By that time a BH has been
formed. The two-sided arrow on the upper left corresponds in this
panel to a length of very approximately $0.1 R_\odot$. }
      \label{fig:schem2}
\end{figure}

We define a region of interest bound by a radius $r_u$.
The mass inside this region just before core-collapse is $M_0(r_u)$,
and the gravitational potential is $\Phi_0(r_u)$.
The mass accreted by the NS is 
$M_\mathrm{a} \equiv \beta M_0\left(r_u\right)
= M_0\left(r_u\right) - M_\nu - \eta_e\left(M_0\left(r_u\right)-M_\mathrm{Fe}\right)$,
where $M_\nu$ is the gravitational mass loss by energetic neutrino emission
($M_\nu \ga 0.2 M_\odot$; \citealt{LovegroveWoosley2013}; see also \citealt{Nadezhin1980,GoldmanNussinov1993}),
$M_\mathrm{Fe}$ is the iron-core mass which is assumed to form the NS before jets are launched,
and $\eta_e$ is the efficiency of the interaction between the jets and the core material.
The overall reduction in gravitational potential at $r_u$ is
\begin{equation}
   \Phi\left(r_u\right) = \beta \Phi_0\left(r_u\right).
 \label{eq:phiru}
\end{equation}
We take the gas to locally obey the virial theorem,
such that the specific pre-explosion internal energy of this gas is
\begin{equation}
e_i = - \Phi_0(r_u)/[3(\gamma-1)],
 \label{eq:eru}
\end{equation}
where $\gamma$ is the adiabatic index.
The new specific energy of the gas after mass removal by jets,
$e(r_u) \simeq e + \Phi(r_u)$, will be
\begin{equation}
   e(r_u) \simeq  - \left[ \frac{1}{3(\gamma-1)} - \beta \right] \Phi_0(r_u) .
   \label{eq:enew}
\end{equation}
If the quantity in the square bracket is positive,
the matter initially becomes unbound.
This condition reads
\begin{equation}
    \beta < 0.83 \left( \frac{\gamma-1}{0.4} \right)^{-1} .
   \label{eq:beta}
\end{equation}

For the purpose of demonstrating our theory we examine two models of rotating massive stars constructed with MESA 
\citep{Paxton2011,Paxton2013,Paxton2015} all the way to core-collapse.
The models have a zero-age main sequence rotation of
$\Omega_\zams=0.55\Omega_\mathrm{crit}$,
where $\Omega_\mathrm{crit}$ is the break-up rotation,
which corresponds to a main-sequence surface rotation of $360\kms$.
Table \ref{table:models} shows several important properties in our models of rotating stars.
Figure \ref{fig:jratio} shows that material around the iron-core in our models
has high enough specific angular momentum to form a rotationally supported structure around the proto-NS,
which we assume launches jets
(the requirement for specific angular momentum above Keplerian may be an exaggeration,
if jets can form from a sub-Keplerian flow, e.g., \citealt{SchreierSoker2016,Soker2016}).
Further details of our models are presented in Appendix \ref{app:mesa}.
Let us now consider the case of a rotating stellar model,
with a zero-age main sequence mass of $M_\zams = 12 M_\odot$
and metallicity of $Z=0.014$.
\begin{deluxetable*}{llllllllll}
\tablecolumns{10}
\tablewidth{0pt}
\tablecaption{Properties of stellar models}
\tablehead{
    \colhead{$M_\mathrm{initial}$} & \colhead{$M_\mathrm{final}$} & \colhead{$M_\mathrm{Fe}$} & \colhead{$v_\zams$} & \colhead{$J_\mathrm{initial}$} & \colhead{$J_\mathrm{final}$} & \colhead{$J_\mathrm{Fe}$} & \colhead{$j_\mathrm{Fe}$} & \colhead{$j_\mathrm{Fe,eq}$} & \colhead{$\Omega_\mathrm{Fe}$} \\
    \colhead{[$M_\odot$]} & \colhead{[$M_\odot$]} & \colhead{[$M_\odot$]} & \colhead{[$\kms$]} & \colhead{[\Jcgs]} & \colhead{[\Jcgs]} & \colhead{[\Jcgs]} & \colhead{[\jcgs]} & \colhead{[\jcgs]} & \colhead{[$\rm{rad}~\rm{s}^{-1}$]}
}
\startdata
	$12$ & $8.62$ & $1.44$ & $361$ & $1.79\times 10^{52}$ & $1.75\times 10^{51}$ & $4.74\times 10^{49}$ & $1.65\times 10^{16}$ & $1.14\times 10^{17}$ & $5.53$ \\
	$54$ & $20.5$ & $1.96$ & $359$ & $2.19\times 10^{53}$ & $7.99\times 10^{51}$ & $6.90\times 10^{49}$ & $1.77\times 10^{16}$ & $7.84\times 10^{16}$ & $2.27$ \\
\enddata
\tabletypesize{\footnotesize}
\tablecomments{The presented properties are from left to right:
initial mass ($M_\mathrm{initial}$),
final mass ($M_\mathrm{final}$),
iron core mass ($M_\mathrm{Fe}$),
zero-age main sequence surface rotation ($v_\zams$),
initial total angular momentum ($J_\mathrm{initial}$),
final total angular momentum ($J_\mathrm{final}$),
iron core total angular momentum ($J_\mathrm{Fe}$),
iron core average specific angular momentum ($j_\mathrm{Fe}$),
outer iron core equatorial specific angular momentum ($j_\mathrm{Fe,eq}$),
iron core angular velocity ($\Omega_\mathrm{Fe}$).}
\label{table:models}
\end{deluxetable*}
\ifmnras
\begin{figure}
\else
\begin{figure}[ht!]
\fi
   \centering
   \ifmnras
   \includegraphics*[scale=0.53]{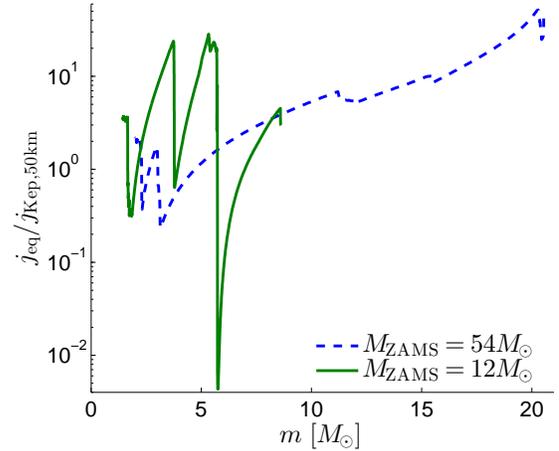} \\
   \else
   \includegraphics*[scale=0.54]{jeqj50ratioModels.eps} \\
   \fi
\caption{The ratio between the equatorial specific angular momentum, and the specific angular momentum for a Keplerian orbit around the proto-NS, as function of mass coordinate for our stellar models. The proto-NS is assumed to have a radius of $50\km$, and mass equal to the pre-collapse iron-core mass (see Table \ref{table:models}). The equatorial specific angular momentum is higher by $50\%$ than the shell average, for the shellular approximation of constant angular velocity in a shell.}
      \label{fig:jratio}
\end{figure}

We define the region of interest as the inner boundary of the helium shell.
In the pre-collapse red supergiant (RSG) star the total mass is $8.6 M_\odot$, and the
mass inner to the helium region is $M_0(r_u) = 3.77 M_\odot$,
with $r_u = 8.2 \times 10^{4} \km$.
The well-collimated jets reach that region with a moderate slow-down
(eq. 6 in \citealt{PapishSoker2011}),
hence in a time of about 1 to 3 seconds post-bounce.
By the time the jets reach $r_u$, the NS is assumed to have a gravitational mass of $1.24 M_\odot$
(according to the iron-core mass of $M_\mathrm{Fe}\approx1.44M_\odot$, and
approximately $0.2 M_\odot$ lost in neutrino emission).
Out of the remaining mass,
$M_0(r_u)-1.44 M_\odot = 2.33 M_\odot$,
we assume that a fraction of about a third, $0.78 M_\odot$, is ejected by jets,
and $1.55 M_\odot$ is accreted in this demonstrative example.
For these values, $\beta = 0.74 $, i.e.,
$\Phi\left(r_u\right) = \left(\left(1.24+1.55\right)/3.77\right) \Phi_0(r_u) = 0.74 \Phi_0(r_u)$.
In the hot core the value of the adiabatic index is $\gamma<5/3$,
with $\gamma\left(r_u\right)=1.4518$ in our example,
so that $\beta$ slightly exceeds the condition in equation (\ref{eq:beta}),
and the material at $r_u$ initially stays bound following the jet activity.
A slight increase (such as from $1/3$ to $0.4$) of the jet interaction efficiency, $\eta_e$,
will decrease $\beta$ and leave the material unbound.
We note that equation (\ref{eq:beta})
is satisfied in most of the outer parts of the star for $\eta_e=0.4$,
while for $\eta_e=0.1$ this condition is not met almost anywhere.
The efficiency of the mass removal by jets will have to be studied in
dedicated hydrodynamical simulations 
(similar to those of, e.g., \citealt{Burrowsetal2007,Couch2009,Lazzati2012};
we note that \citealt{PapishSoker2014b} find a high efficiency, but for the jittering jets scenario).
Figure \ref{fig:efficiency} illustrates the efficiency dependence
we claim for the outcome of the core-collapse.
\ifmnras
\begin{figure}
\else
\begin{figure}[ht!]
\fi
   \centering
   \ifmnras
   \includegraphics*[scale=0.53]{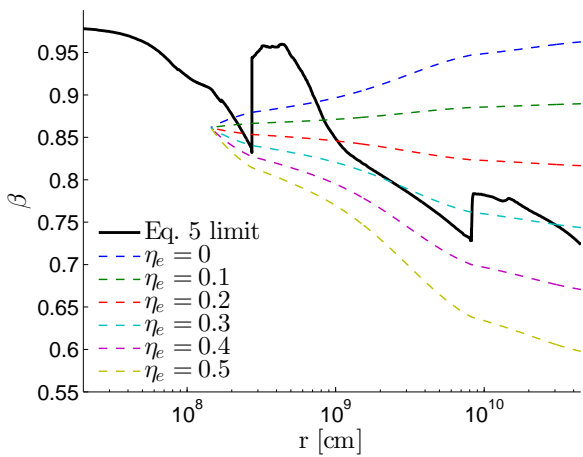} \\
   \else
   \includegraphics*[scale=0.54]{efficiency12.eps} \\
   \fi
\caption{Presentation of the condition for material to become energetically unbound according to Eq. 
(\ref{eq:beta}).
The solid black line shows the right-hand side of Eq. (\ref{eq:beta})
for our pre-collapse stellar model with $M_\zams = 12 M_\odot$.
Colourful dashed lines show the value of $\beta$,
the factor by which the gravitational potential is reduced after polar mass ejection by jets,
for different assumed jet efficiencies of expelling matter ($\eta_e$).
The branching point is at the pre-collapse radius of the mass coordinate
which we assume is accreted before mass begins being expelled by jets.
The value of $\beta<1$ at that point is due to the assumed mass-loss
from neutrino emission in the first several post-bounce seconds \citep{LovegroveWoosley2013}.
It can be seen that for high efficiency ($\eta_e \ga 0.4$) the condition of Eq. (\ref{eq:beta})
is satisfied everywhere -- this is the case of high efficiency of the JFM.
Namely, the accretion onto the compact object is halted relatively early-on
and a regular supernova explosion takes place (and not a SESN).
For low efficiency the gravitational potential is not reduced significantly
and material accretion continues for a considerable duration,
powering an energetic outflow.
$\eta_e=0$ corresponds to no jet activity.
We note again that the condition of Eq. (\ref{eq:beta})
is approximate and the dynamics will have to be studied with hydrodynamic simulations.}
      \label{fig:efficiency}
\end{figure}

The above values suggest that in some cases there is a radius $r_c$
within the core where $e(r_c)=0$.
Matter in the region just inward to $r_c$
will at first move outward to a relatively large distance,
but then will fall back to be later accreted onto the NS or the BH.
If this `sluggish' gas is massive enough,
jets with significant energy can power the explosion for a long time
(or power the post-explosion ejecta).
We note that even if the matter has an initial positive energy,
it collides with regions above it,
and transfers energy to outer parts.
So even mass with an initial positive energy might lose energy and eventually might fall back.
Still, some material may stay unbound and not fall back.
This will quantitatively affect the amount of matter accreted,
and thus the power of the jets and the remnant mass.
The fate of this slow equatorial flow is further discussed in section \ref{subsec:slowflow}.

We can estimate the explosion energy of the example studied here in the frame of the JFM.
For the values used in the representative example above,
accretion of material up to the base of the helium layer (not including the ejected polar gas)
will produce a BH with a mass of $M_\mathrm{BH} \approx 2.8 M_\odot$.
Using equation (\ref{eq:Etot}),
we can get a total explosion energy of $E_\mathrm{tot} \approx 1.6 \times 10^{52} \erg$,
under the assumption that $\approx 10\%$
of the accreted mass is launched into the jets at a speed of about $10^5 \km \s^{-1}$.
If accretion from the helium shell continues in a similar manner to the preceding phase,
a mass of $\approx 1.3 M_\odot$  will be accreted after a BH was formed.
We can take the energy carried by the jets formed from the mass accreted after a BH was formed to be
about $10 \%$ of the rest energy, as is usually assumed for jets launched by BHs.
We then derive the energy carried by the jets to be
$\approx 2.5 \times 10^{53} \erg$. 
A higher efficiency of jets will give a lower explosion energy and leave a lower mass remnant,
while a lower efficiency will do the opposite.
This is an illustration of a SESNe which can be an outcome of a rapidly-rotating progenitor star
coupled with the inefficient feedback process.

\subsection{A slow equatorial outflow}
\label{subsec:slowflow}

Gas from near the equatorial plane that becomes unbound
because of the gas removal from polar directions
forms a slow equatorial outflow.
Pressure from the jet-inflated bubble might accelerate it somewhat and compress it,
but it will stay a slow equatorial outflow.
Based on equation (\ref{eq:enew}) the positive energy is a fraction of the initial gravitational energy,
hence its speed will be a fraction of the escape speed from the relevant radius
$r_k \approx 0.1-1R_\odot$,
defined to be the initial radius of the material that later falls back onto the central BH
(see Fig. \ref{fig:schem2}).
For the considered masses $M(r_k) \approx 10 M_\odot$,
the escape speed is $v_\mathrm{esc} \approx 2000-6000 \kms$.
We therefore crudely expect the slow equatorial outflow to be somewhat slower than the escape velocity,
with a typical speed of
$v_\mathrm{eq} \approx 1000 \kms$
(but the range can be of $v_\mathrm{eq} \approx 500-3000 \kms$).
The mass in this outflow is $M_\mathrm{eq} \ga 1 M_\odot$.
Such a slow equatorial outflow is a prediction of the inefficient JFM scenario for SESNe.

The expulsion of the hydrogen-rich envelope as a result of the
decrease of the NS and then BH mass by neutrino cooling was studied by
\cite{Nadezhin1980} and \cite{LovegroveWoosley2013} in a spherical failed SN.
We here differ in several points.
($i$) We consider a scenario of a successful jet-driven CCSN explosion, not with a failed SN.
($ii$) We consider the effect of the core mass reduction by jets in a non-spherical flow,
which we expect to be more significant than mass loss by neutrino cooling.
($iii$) We study the effect on the outer parts of the core, rather than only on the hydrogen-rich envelope.
($iv$) We are not interested in the expulsion of gas,
but rather in the process of `lifting' core gas to large radii,
from where the `sluggish' gas falls back and forms a late accretion disk.
Still, some gas is expelled along the equator as an interesting side-effect.


\subsection{A large accretion disk}
\label{subsec:largedisk}

We now turn to
the effect of rotation mainly on core layers further out,
that are accreted at later times with respect to the example of subsection \ref{subsec:inefficient}.
We consider some point $r_0$ in the outer part of the helium shell.
The rotation velocity there is
$\Omega\left(r_0\right)=\xi_0\Omega_\mathrm{Kep}$,
where $\Omega_\mathrm{Kep}$ is the local break-up rotation velocity.
We assume that after mass removal from the core by jets,
regions near the equatorial plane but further out,
$r \gg r_0$, leave the star
because of the decrease in the central mass, and hence in gravity.
Material near $r_0$ stays bound, but is temporarily lifted out.
This material might initially have a positive energy,
becoming in principle unbound, but losing energy due to interactions with outer regions.
When the mass falls back it forms a Keplerian disk at a radius of
\begin{equation}
 r_d =  r_0 \xi^2_0 \beta^{-1}
 \label{eq:rd}
\end{equation}
where $\beta$ was defined in equation (\ref{eq:phiru}).
\cite{Kumaretal2008} considered already the formation of an
accretion disk from fall-back gas,
but they did not consider the removal of gas from the polar directions.

The accretion time lasts longer than the viscous timescale
for the accreted mass to lose its angular momentum.
The viscous timescale is
\begin{equation}
\begin{split}
& t_\mathrm{visc}  \simeq \frac{r_d^2}{\nu} \simeq 5\times 10^4
\left(\frac{\alpha}{0.1}\right)^{-1}
\left(\frac{H}{0.1r_d}\right)^{-1}
\left(\frac{C_s}{0.1v_\phi}\right)^{-1}
     \\   &   \times
     \left(\frac{M_a}{10~M_{\odot}}\right)^{-1/2}
   \left(\frac{r_0}{R_{\odot}}\right)^{3/2}
\left(\frac{\xi_0}{0.25}\right)^{3}
\left(\frac{\beta}{0.3}\right)^{-3/2}
 \s, 
\label{eq:tvisc1}
\end{split}
\end{equation}
where $\nu=\alpha C_s H$ is the viscosity of the disk,
$H$ is the thickness of the disk,
$C_s$ is the sound speed,
$\alpha$ is the disk viscosity parameter,
$v_\phi$ is the Keplerian velocity,
and $M_a$ is the mass inside $r_0$ after material was removed by jets.
For these parameters the viscous to Keplerian times ratio is $\chi \equiv t_\mathrm{visc}/t_\mathrm{Kep} \simeq 15$.

We return to our models of rotating massive stars
(see Table \ref{table:models} and Fig. \ref{fig:jratio};
Appendix \ref{app:mesa} for further details).
The evolution of the angular momentum in our stellar models
\emph{excludes} the effect of the Spruit-Tayler dynamo \citep{Spruit2002},
hence reducing angular momentum loss in the stellar wind and
resulting in high core rotation.
The specific angular momentum of the iron-core is above the break-up value for a NS,
so the angular momentum of the accreted gas
must be lost by some mechanism after core-collapse,
perhaps through an accretion disk accompanied by jets.
We expect that realistically likewise high rotation (or higher)
will result from specific types of binary interaction (e.g., \citealt{Izzardetal2004, Podsiadlowskietal2004, FryerHeger2005, Cantiello2007, Yoonetal2010, deMink2013}).
Although the effect of the Spruit-Tayler dynamo is at issue \citep{Braithwaite2006,Zahn2007,Cantiello2014},
we use the single-star approach for simplicity,
and it is sufficient for the purpose of our presentation.

We look in our stellar models at the coordinate of maximal specific angular momentum,
which is a good indicator for the duration of a late accretion episode.
The parameters at that point for the $M_\zams=12M_\odot$ model,
that evolves to a pre-collapse RSG with a total mass of $M = 8.6M_\odot$,
are $r_0=0.64R_\odot$, $M_0=5.3M_\odot$, and $\xi_0=0.16$.
For the $M_\zams=54M_\odot$ model,
that becomes a pre-collapse Wolf-Rayet (WR) star with a total mass of $M=20.54M_\odot$,
we find $r_0=0.28R_\odot$, $M_0=20.27M_\odot$ and $\xi_0=0.26$.
The coordinate of maximal specific angular momentum is near the outer part of the helium shell
for the RSG model,
and near the outer part of the oxygen-neon shell for the WR.
Applying equation (\ref{eq:rd}) with $\beta=0.3$,
the above parameters give accretion disks with radii of
$r_d\left(M_\zams=12M_\odot\right) = 3.6\times10^4\km \simeq 0.05 R_\odot$
and
$r_d\left(M_\zams=54M_\odot\right) = 4.3\times10^4\km \simeq 0.06  R_\odot$,
with
$t_\mathrm{visc}\left(M_\zams=12M_\odot\right)\approx 8.6\times 10^3 \s \approx 2.5 \h$
and
$t_\mathrm{visc}\left(M_\zams=54M_\odot\right)\approx 5.7\times 10^3 \s \approx 1.5 \h$.

We can see that in a rapidly rotating core,
a large accretion disk might supply gas to the central object,
now a BH, for hours.
By large accretion disk we refer to a disk that is about $10 - 10^4$
times larger than the radius of the last stable orbit around the BH.

\subsection{A very long accretion phase}
\label{subsec:longaccretion}

To form a thin accretion disk the gas should cool down.
In the distances studied here,
$r_d \approx 0.01-1 R_\odot$
(hatched region in Fig. \ref{fig:schem2}),
the cooling is via photon diffusion,
and not neutrino cooling.
The thermal timescale for a pre-collapse shell of mass $M_s$
is $t_\mathrm{th} \approx G M_0 M_s / r_0 L_0$,
where $L_0$ is the luminosity at $r=r_0$.
For the models considered in section \ref{subsec:largedisk},
taking $M_s=1M_\odot$ gives
$t_\mathrm{th} \left(M_\zams=12M_\odot\right) \approx 900 \yr$
and
$t_\mathrm{th} \left(M_\zams=54M_\odot\right) \approx 1200 \yr$
in the close vicinity of $r_0$.
The diffusion time out of a disk of a size
$r_d\approx 0.1 r_0$ is $t_\gamma\approx 10 \yr$.
These very long timescales have the following implication.
Once the high angular momentum gas forms an accretion disk at
$r_d \approx 0.01-1 R_\odot$,
the gas in the disk requires
a time much longer than the viscous time given in
equation (\ref{eq:tvisc1})
to get rid of its thermal energy.
The extra energy allows the disk to spread outward while some gas is accreted.
When a fraction of $\approx 0.1$ of the disk mass has spread to $\approx 10 R_\odot$,
for example,
the photon diffusion time is about equal to the viscosity time.

This order of magnitude consideration suggests that a very-rapidly
rotating pre-collapse core might lead under rare conditions to the formation of
an accretion disk around the central remnant BH that spreads out to $\approx 10 R_\odot$.
A very-late accretion of, say,
$M_\mathrm{vl}\approx 0.01 M_\odot$
of that gas over the viscous time of
$t_\mathrm{vl} \simeq 1 \yr$,
might under extreme conditions power a SESN with
$\dot E_\mathrm{vl-jet} \approx 0.1 M_\mathrm{vl} c^2/t_\mathrm{vl} \approx 5 \times 10^{43} \erg \s^{-1}$
a year after explosion.
However, in most of these rare cases we expect the accretion to last for up to a week,
or even only a day (for $\alpha \simeq 0.5$),
as the timescale given in equation (\ref{eq:tvisc1}).
In the majority of cases the core will not rotate so rapidly,
and a more typical accretion time will be about hundreds to
thousands of seconds as derived in section \ref{subsec:largedisk}.
3D hydrodynamic simulations are definitely required to explore this flow.

The poorly known pre-collapse angular velocity distribution in the
core is a key quantity that determines the outcome.
Different pre-collapse conditions may lead to very diverse accretion times,
which may help consolidate the large span of LGRB times (e.g., \citealt{Zhang2014}).
In our proposed scenario the so-called ultra-long GRBs may be qualitatively
the outcome of the same type of explosion and late accretion process as in LGRBs,
but with extreme rotation,
contrary to the suggestion by \cite{Boer2015} of a distinct class of transients.

The jets that are launched days to months after explosion will collide with the ejecta,
and heat it through shock waves.
In the optically thick regime,
up to months after explosion,
the extra energy will reveal itself as a slowly fading super-luminous SN.
During the optically thin phase strong emission lines and X-ray emission,
as in colliding wind of massive stars,
will appear.
The calculation of the exact light-curve and spectrum is beyond the scope of the present paper.

The fall-back process onto a NS has been studied before.
\cite{Chevalier1989} calculated the expected fall-back flow onto a NS in SN~1987A in a spherical geometry.
The fall-back can last for hours to months,
with approximately $0.1 M_\odot$ of fall-back gas.
We here differ from \cite{Chevalier1989} and from \cite{Kumaretal2008}
in discussing a highly non-spherical flow geometry
that is based on jet-driven explosion acting through a feedback process.
This, we argue, can lead to a fall-back mass of $\ga 1 M_\odot$,
an order of magnitude more than in the flow studied by \cite{Chevalier1989}.

\subsection{Connection to gamma-ray bursts}
\label{subsec:grbs}

Jets that propagate through the stellar envelope may be collimated
and can give rise to a LGRB (e.g., \citealt{Brombergetal2014,Bromberg2016}).
The timescales considerations in subsection \ref{subsec:longaccretion}
suggest a possible prolongation of the LGRB duration.
However, this depends on the continues collimation of the jets.

In the first several minutes of the accretion phase part of
the stellar envelope is still intact.
Later, the density along the polar directions is much lower,
and efficient collimation might not take place anymore for very late jets.
The emission of strong gamma rays from highly-relativistic jets may be then inhibited.
In that case,
the jets can emit X-ray radiation and they can collide with the SN ejecta and power it.
Details of these mechanisms will have to be studied in the future.

\subsection{A fast rise to peak magnitude}
\label{subsec:fastrise}

Recent studies find luminous transient events with a rapid rise to peak magnitude
\citep{Drout2014,Arcavietal2016}.
We suggest a process in which
jets might account for this observed phenomenon, as follows.
CCSNe rise to maximum by the diffusion of photons outward.
In the conjectured process, the jets penetrate through inner ejecta layers and stop at outer regions.
The jets heat these regions. As the photosphere moves to inner layers,
it encounters this hot region at an earlier time.
The location of the photosphere in a hot region and at an early time makes
the SN both super-luminous and having a fast rise to maximum.

If our rationale holds,
then the events studied by \cite{Arcavietal2016}
are part of a continuous range of explosion energy from regular CCSNe to extreme SESNe,
and are not a class of a different explosion mechanism.
The efficiency of the JFM is the main factor that varies along this range,
from being very efficient in regular CCSNe to being very inefficient in extreme SESNe
with a kinetic energy of several times $10^{52} \erg$ and up to $10^{53} \erg$.
The fast rise to peak magnitude will be studied in a future paper.

\section{DISCUSSION AND SUMMARY}
\label{sec:summary}

The present study was conducted within the premise that the explosion
mechanism of CCSNe operates through a negative feedback mechanism.
This is supported by the finding that the kinetic energy of most CCSNe
is a few times the binding energy of the core that is ejected in the explosion.
When the explosion mechanism transfers to the core an energy of about its binding energy,
the supply of accreted mass from the core ends, and the explosion terminates.
As the efficiency of energy transfer is never 100\%,
the explosion energy is larger than the binding energy.

We further take the view that the explosion is via a negative jet-feedback mechanism (JFM).
The JFM is thought to operate in many different astrophysical objects:
galaxy formation, cooling flow clusters,
in some cases of the the common envelope evolution,
in the grazing envelope evolution,
and in some intermediate luminosity optical transients
(see Table 1 in \citealt{Kashisoker2016}).
In the case of CCSN explosions,
the JFM is termed the jittering jets mechanism (Fig. \ref{fig:schem1}).
Most efficiently the JFM works when the jets inflate large (`fat') bubbles \citep{Sokeretal2013}.
To inflate `fat bubbles',
a relative transverse velocity between the jets and the ambient gas is required.
When the jets maintain a constant and stable direction,
and the ambient gas, in the present case the core,
has no large scale transverse velocity relative to the jets,
the JFM is less efficient.
This is the case studied here.

When the pre-collapse core is rapidly rotating an accretion disk with a constant axis is formed.
The jets that are launched by the disk maintain a constant axis,
and mass in the equatorial plane is not removed efficiently.
However, the removal of a large fraction of the core mass from regions away
from the equatorial plane reduces the gravitational potential,
and mass from near the equatorial plane moves outward (Fig. \ref{fig:schem2}).
Part of this mass might fall back and form an accretion disk at
$r_d \approx 0.01-1 R_\odot$
(for other fall-back scenarios see, e.g., \citealt{Kumaretal2008,KashiyamaQuataert2015}).
The accretion timescale from this disk might be longer than the viscous timescale
(eq. \ref{eq:tvisc1}), up to days and weeks.
However, in most cases the accretion phase will be much shorter.
In any case, we suggest in this study that the prolonged accretion phase of
$\ga 1 M_\odot$ accounts for very energetic SESNe,
such as the very rare CCSNe cases
ASASSN-15lh \citep{Dongetal2015} and iPTF13ehe \citep{Wangetal2015}.

Synthesis of nickel-56 is an unresolved issue of jet-driven CCSN explosions.
\cite{Nishimura2015}
obtained $\mathrm{few}\times0.01 M_\odot$
of synthesized nickel-56 in their study of jet-driven explosions.
\cite{Milosavljevic2012} suggest that enough nickel is produced 
in a collapsar accretion flow, powering jets,
to explain light curves of supernovae
associated with LGRBs
(see also \citealt{MacFadyenWoosley1999}).
The synthesis of nickel-56 (as well as calculation of light-curves and spectra)
will be studied in future work.

Three-dimensional numerical simulations that include the radius from $10 \km$ to $10 R_\odot$
are required to explore the flow structure depicted in Fig. \ref{fig:schem2},
and in particular to find the duration of the jets-launching phase
as a function of the pre-collapse core angular velocity profile.
As well, such simulations will explore the properties of the slow equatorial outflow.
These are extremely demanding simulations,
above and beyond the resources of our group.

In relation to GRBs,
we bring here the very recent study of GRB~130831A conducted by \cite{DePasqualeetal2015}.
This GRB was associated with the energetic SN~2013fu,
having a kinetic energy of $E_\mathrm{k} \simeq 1.9 \times 10^{52} \erg$ \citep{Canoetal2014},
and had an X-ray emission source that lasted for $\approx 1~$day.
\cite{DePasqualeetal2015} stated that accretion of fall-back matter onto the central BH for
$t_\mathrm{fb}\approx1~\days$ requires a large fraction of the envelope mass.
In the flow geometry studied here, on the other hand,
only a small fraction of the core mass is required to fall-back,
as only material from near the equatorial plane falls back.
\cite{DePasqualeetal2015} favor a magnetar as the
explanation for the long lasting X-ray source.
We note that the fall-back accretion onto a BH and the launching of jets has an advantage,
as it nicely connects the long-lasting engine to the engine of all CCSNe with $E_\mathrm{tot} \ga 0.5 \foe$.

The connection between LGRBs and SESNe is an issue of ongoing study
(e.g., \citealt{Fruchter2006,Woosley2006,Bissaldi2007,Modjaz2008,Modjaz2011}).
We suggest that LGRBs and SESNe are just extreme cases of the JFM,
where the feedback efficiency is very low,
hence the explosion on average is {\it more energetic}
(\citealt{Modjazetal2016} claim that Type Ic SNe with GRBs are more energetic on average than those without GRBs).

According to the model for SESNe proposed in the present study,
\emph{the main explosion parameter that distinguishes between the
different cases of CCSN explosions is the efficiency of the jet-feedback mechanism.}
Of course, this efficiency in turn depends on the pre-collapse core mass,
envelope mass, core convection, and above all on the pre-collapse angular momentum profile in the core.
This angular momentum requires specific types of binary evolution,
and the formation of jets from the accretion disk around the newly formed NS or BH
has not been demonstrated yet.
These are future challenges for the JFM.

\section*{Acknowledgments}

We thank an anonymous referee for useful comments.
N.S. is supported by the Charles Wolfson Academic Chair.

\appendix
\section{Implication of angular momentum required by the CITE mechanism}
\label{app:cite}

In the collapse-induced thermonuclear explosion (CITE) mechanism
\citep{Burbidgeetal1957, KushnirKatz2014, Kushnir2015a, BlumKushnir2016}
the source of the explosion energy is nuclear burning rather than gravitational energy.
The collapse-induced thermonuclear explosion (CITE)
mechanism for CCSNe presented by \cite{Kushnir2015a} requires a rapidly rotating pre-collapse core.
Although presented as `slow' rotation,
our straightforward analysis of the specific angular momentum profile in the stellar model used by
\cite{Kushnir2015a} shows that a large accretion disk is formed
(our claim for disk formation was confirmed by \citealt{BlumKushnir2016}).
The jets that are expected to be launched by this accretion
disk will dwarf the energy released by the thermonuclear burning.

In large parts of the core the specific angular momentum required by the CITE mechanism
is much higher than that of the break-up rotation velocity of a proto-NS (or a BH).
This leads to the formation of a large accretion disk round the newly born Ns.
We take the angular velocity profile as used by \cite{Kushnir2015a} and calculate the
radius at which the collapsing gas forms an accretion disk around the newly born NS,
as a function of its pre-collapse radius.
The results are presented in Figure \ref{fig:kushnir}.
We see, for example, that the gas starting at a mass coordinate of about $2 M_\odot$
and a radius of about $1.5 \times 10^4 \km$,
attains its Keplerian velocity at a radius of about $800 \km$ around the newly born NS.
Such a large accretion disk is likely to form energetic jets, which are expected to carry
more energy than that obtained from the thermonuclear burning of the He-O layer.
A release of thermonuclear explosion energy from a mixed He-O layer might quench further accretion.
Depending on the total mass of the mixed He-O,
and the remaining mass which will not be accreted,
the CITE mechanism might actually \emph{reduce} the energy output of the SN,
if the SN is powered by jets as we argue here.
\ifmnras
\begin{figure}
\else
\begin{figure}[ht!]
\fi
   \centering
   \ifmnras
    \includegraphics[scale=0.55]{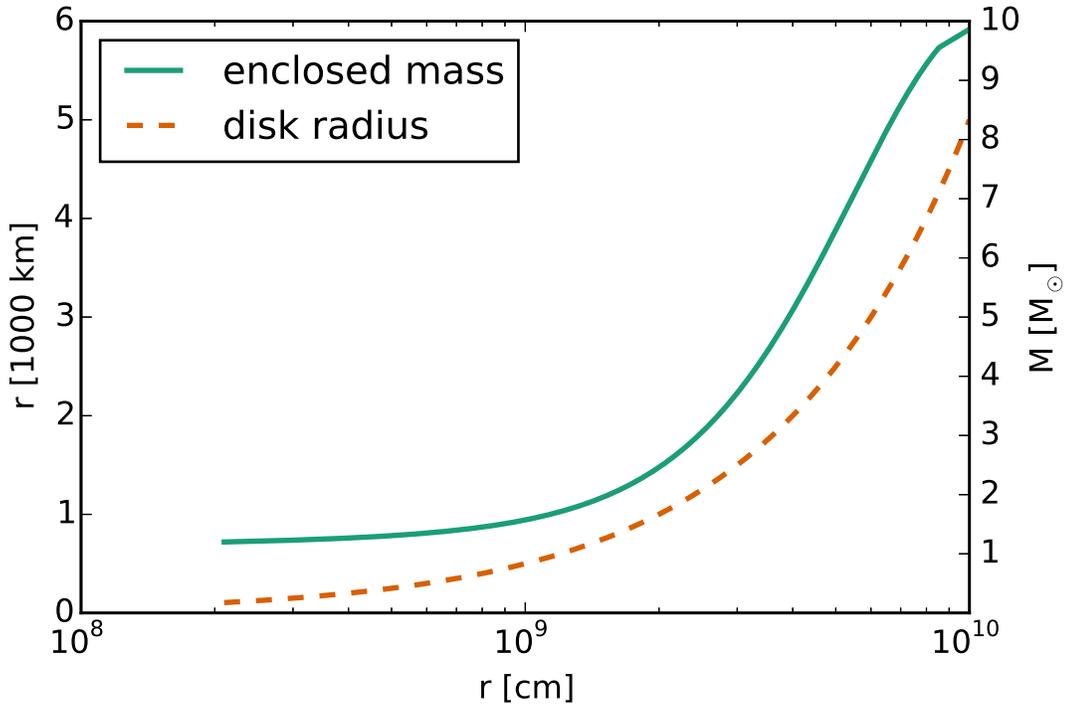} \\
   \else
    \includegraphics{kushnir_plot.eps} \\
   \fi
    \caption{The enclosed mass (upper line) and the radius (lower line)
    at which the accreted gas attains its Keplerian velocity around
    the newly born NS as a function of its pre-collapse radius.
    The pre-collapse angular velocity profile in the core
    is taken form the model presented in Fig. 1 of \citet{Kushnir2015a},
    but without a lower cutoff of the angular momentum.
    The in-falling material has enough angular momentum
    to create a large accretion disk around the newly born NS (or BH).
    This accretion disk is likely to launch very energetic jets that expel the envelope.}
      \label{fig:kushnir}
\end{figure}

\section{Models of rotating massive stars evolved with MESA}
\label{app:mesa}

We constructed a set of stellar models using
Modules for Experiments in Stellar Astrophysics
(MESA version 7624; \citealt{Paxton2011,Paxton2013,Paxton2015}),
with an initial mass between $M_\zams=12$ and $90 M_\odot$,
and initial rotation in the range $0.1 \le \Omega / \Omega_\mathrm{crit} \le 0.9$.
All models have a metallicity of $Z=0.014$.
Rotation is implemented in MESA using the `shellular approximation' \citep{Meynet1997},
where the angular velocity $\Omega$ is assumed to be constant over isobars.
All models were evolved well into the silicon shell burning stage,
and have an iron-core mass of $M_\mathrm{Fe}\ga 1.4M_{\odot}$.
Convection is treated according to the Mixing-Length Theory with $\alpha_\mathrm{MLT}=1.5$.
Semiconvective mixing \citep{Langer1983,Langer1991} is employed with $\alpha_\mathrm{sc}=0.1$.
Exponential convective overshooting is applied as in \cite{Herwig2000},
with $f=0.016$ (the fraction of the pressure scale height for the decay scale).
Mass loss was treated according to the so-called `Dutch' scheme \citep{Nugis2000,Vink2001,Glebbeek2009},
with rotational enhancement \citep{Heger2000,Maeder2000}.
Rotationally-induced instabilities and convection transport angular momentum within the stellar models 
\citep{Paxton2013}.
The Spruit-Tayler dynamo \citep{Spruit2002} is not included,
effectively resulting
in a higher rotational velocity than might be expected for single-stellar evolution.
Late spin-up by a stellar companion
may result in similar or higher core rotation,
as is expected for rapidly rotating CCSN progenitors
(e.g., \citealt{Cantiello2007,deMink2013}).
About half out of several tens of models constructed this way had pre-collapse iron-cores
with angular momentum higher than the break-up value for a NS with equivalent mass.
For the same initial parameters but with the Spruit-Tayler dynamo taken into account,
the number of highly-rotating pre-collapse cores drops to zero
(this distinction has been noted by \citealt{Yoon2015}, and references therein).

Clearly, there are numerous uncertainties in the modeling of massive stars,
specifically regarding rotation, convection and mass-loss which were mentioned above
(\citealt{Langer2012} for a review).
These uncertainties may reflect quantitatively on the jet-feedback mechanism discussed in the present paper.
In future works the evolutionary channels and conditions for the operation of the jet-feedback mechanism
should be studied in detail.
In the present study we are content with self-consistent models of rotating stars
where the angular momentum of the pre-collapse core is high enough for accretion disk formation,
and we present two stellar models out of the constructed set for our demonstrative purposes.
The presented models were taken from the high-end of the resulting iron-core rotation distribution,
with one giant star and one Wolf-Rayet (WR) star chosen.
The initial masses of the models are $M_\zams=12$ and $54 M_\odot$,
with an initial rotation of $\Omega=0.55\Omega_\mathrm{crit}$ for both.
Due to stellar winds the final masses are $8.6M_\odot$ and $20.5M_\odot$, respectively.
The heavier model loses its hydrogen envelope and becomes a WR star,
while the lighter model becomes a red supergiant (RSG).
The detailed composition,
evolution and angular velocity of the stellar models are presented in Figure \ref{fig:models1}.
The models show many similarities to the stellar models with rotation of \cite{Ekstrom2012}
(see also \citealt{Groh2013}).
\ifmnras
\begin{figure*}
\else
\begin{figure}
\fi
\begin{tabular}{cc}
\ifmnras
{\includegraphics*[scale=0.53]{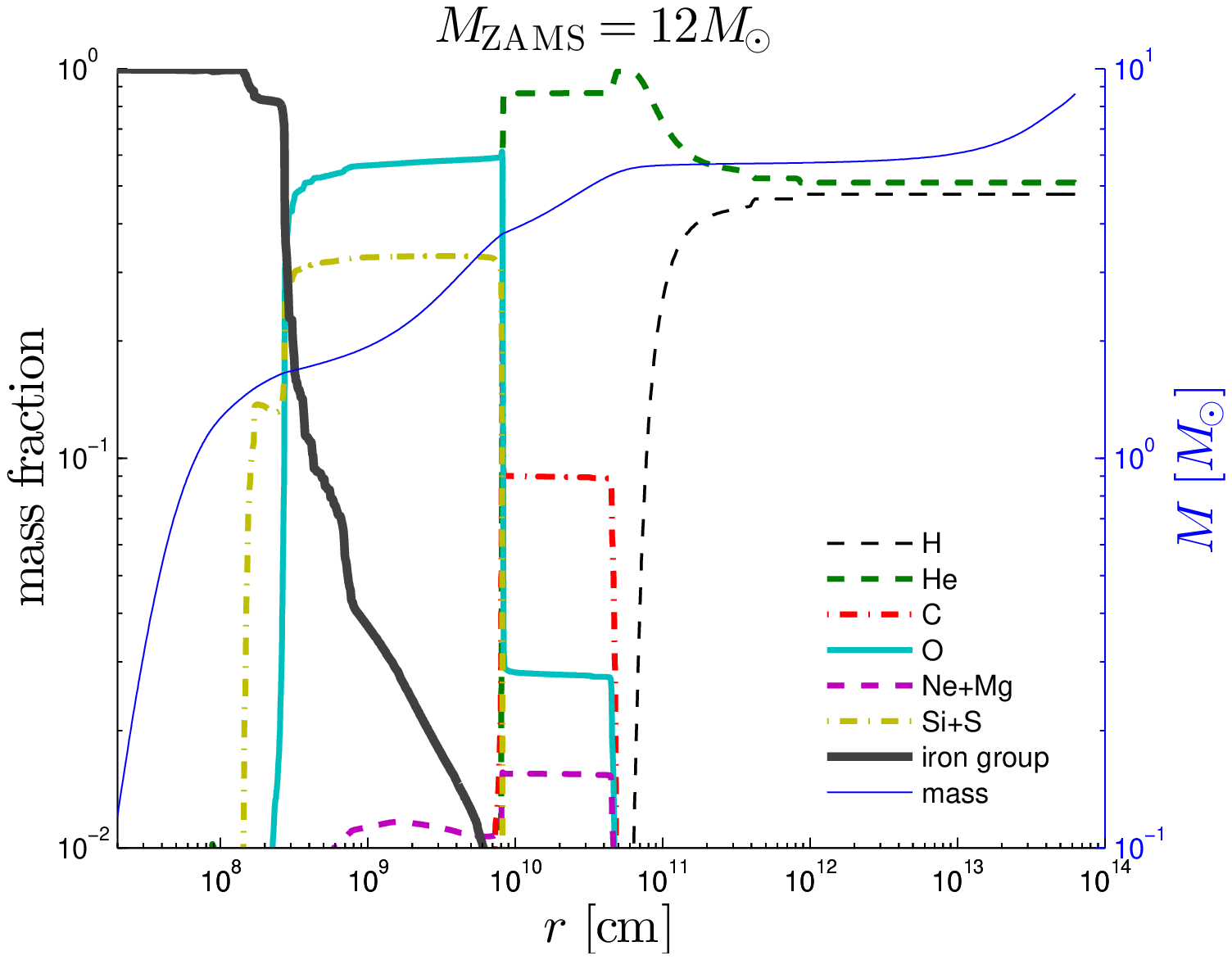}} &
{\includegraphics*[scale=0.53]{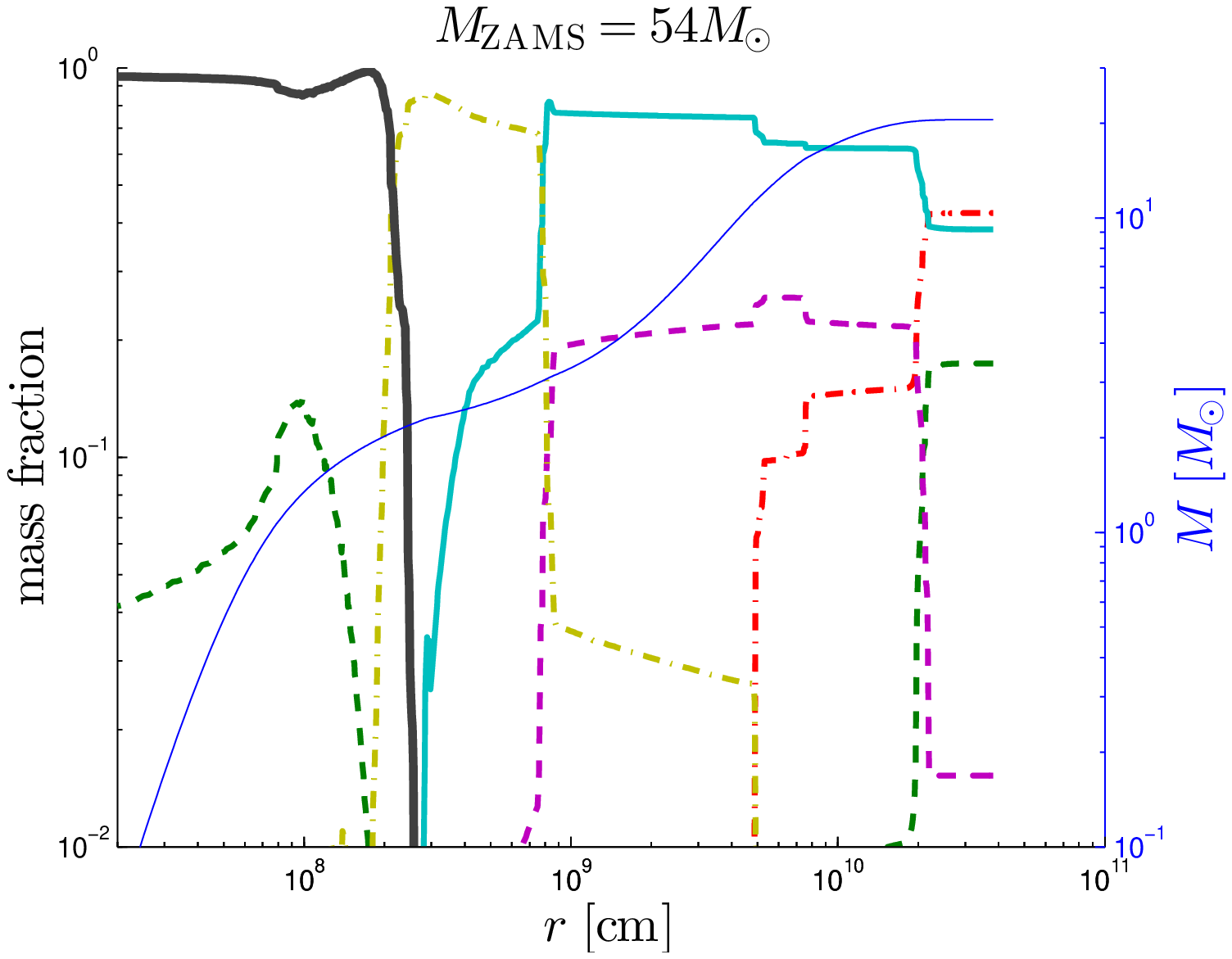}} \\
{\includegraphics*[scale=0.53]{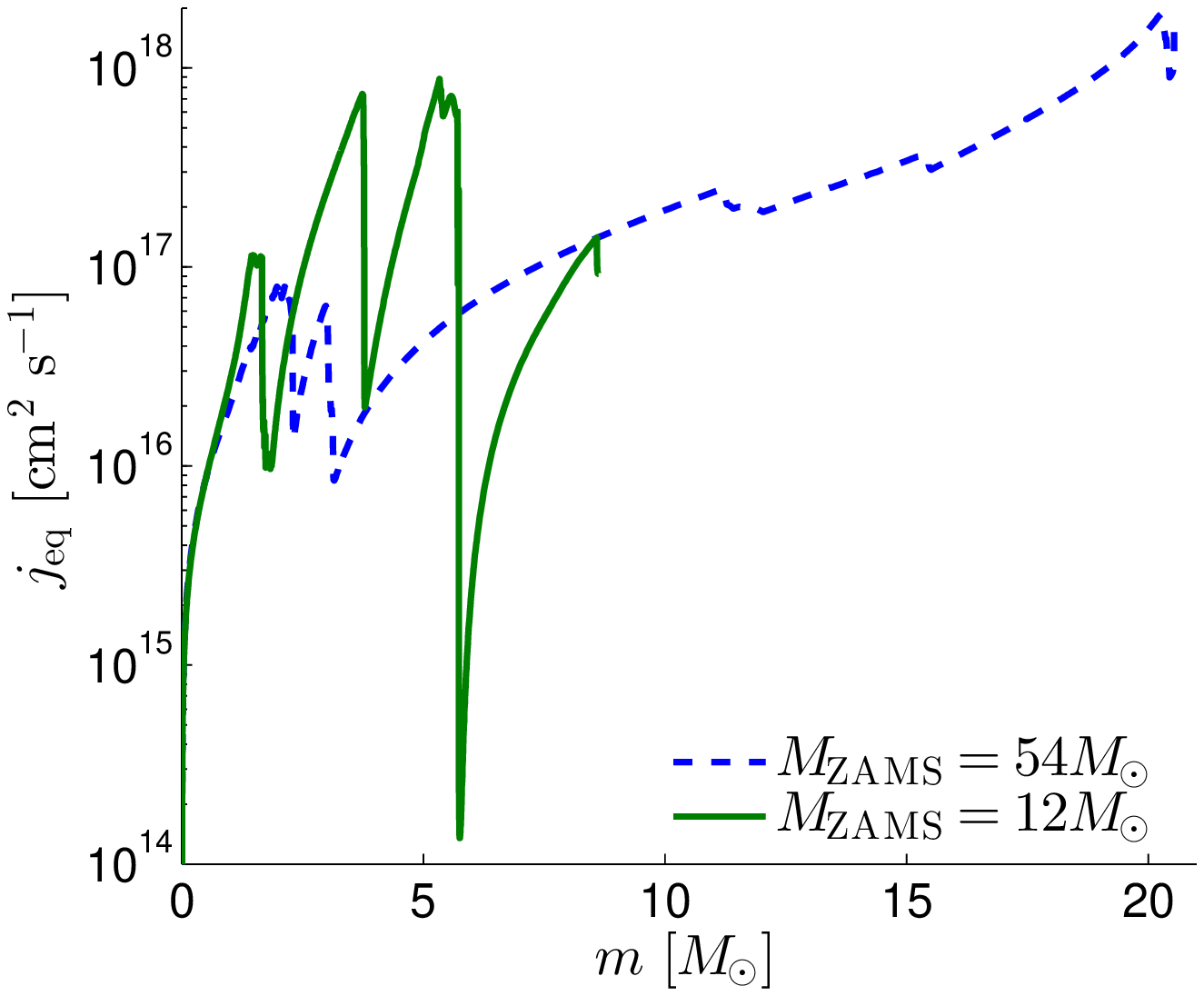}} &
{\includegraphics*[scale=0.53]{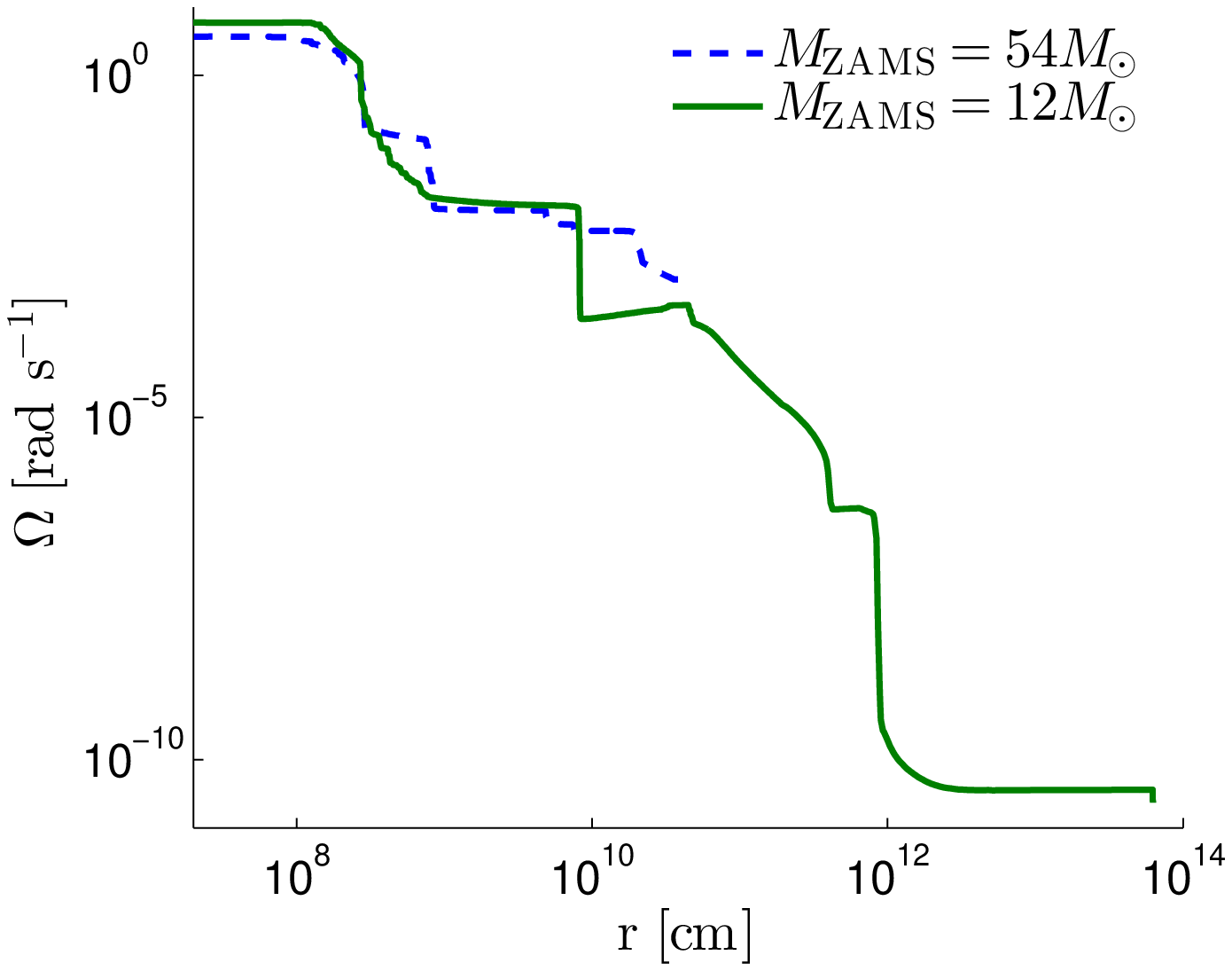}} \\
{\includegraphics*[scale=0.53]{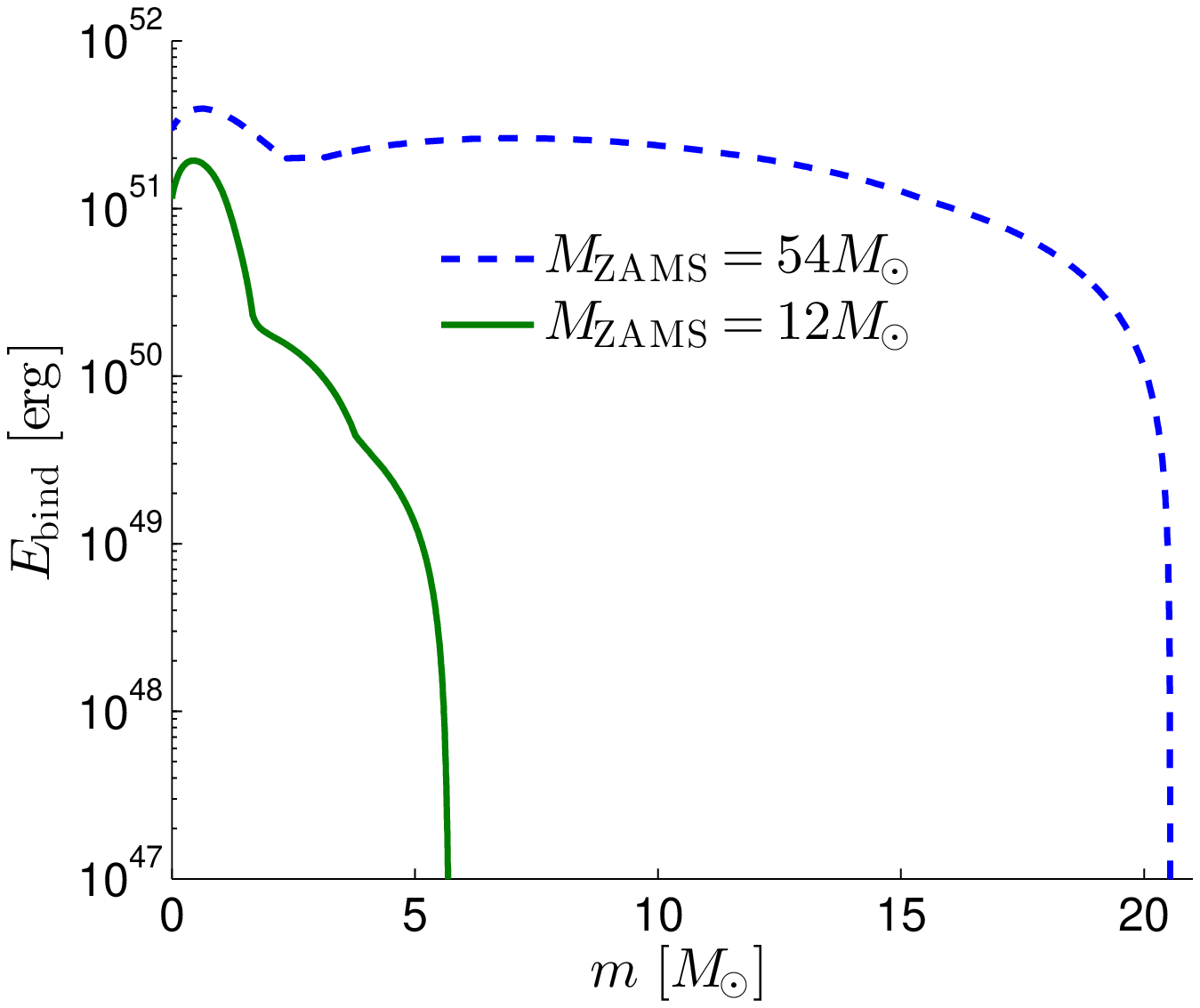}} &
{\includegraphics*[scale=0.53]{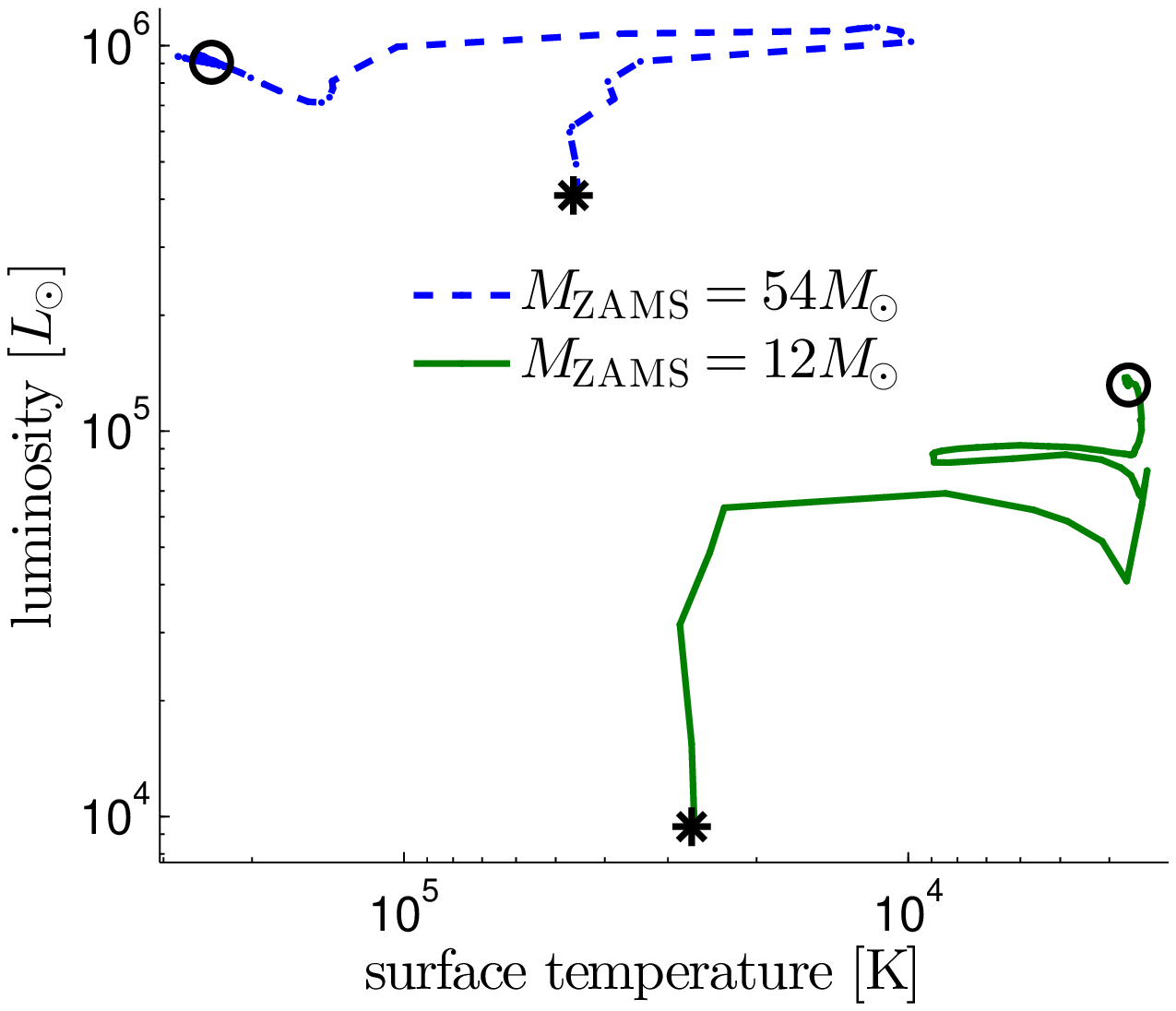}} \\
\else
{\includegraphics*[scale=0.54]{CompoM12Omega055new.eps}} &
{\includegraphics*[scale=0.54]{CompoM54Omega055new.eps}} \\
{\includegraphics*[scale=0.54]{Mov0jeqModels.eps}} &
{\includegraphics*[scale=0.54]{Mov0OmegaModels.eps}} \\
{\includegraphics*[scale=0.54]{Ebind.eps}} &
{\includegraphics*[scale=0.54]{HR.eps}} \\
\fi
\end{tabular}
      \caption{Top row: Composition and mass distribution
      of the stellar models just prior to core-collapse.
      The top-left panel shows a model with $M_\zams=12M_\odot$,
      which has a pre-explosion mass of $M=8.6 M_\odot$,
      photospheric radius of $R=903R_\odot$,
      luminosity of $L=1.3 \times 10^5 L_\odot$
      and effective temperature of $T=3.7\times 10^3 \K$.
      The top-right panel shows a model with $M_\zams=54M_\odot$,
      $M=20.5 M_\odot$,
      $R=0.55 R_\odot$,
      $L=9.1 \times 10^5 L_\odot$
      and $T=2.4\times 10^5 \K$.
      Middle-left: Pre-collapse equatorial specific angular momentum as function of interior mass for the two stellar models.
      Middle-right: Pre-collapse angular velocity as function of interior radius.
      Bottom-left: Gravitational binding energy
      (absolute value of integrating $E_\mathrm{grav}+E_\mathrm{int}$ from $E_\mathrm{bind}=0$ at the surface, inwards).
      Bottom-right: Evolutionary tracks for the two stellar models from the main sequence
      (star symbols) to pre-collapse (circle symbols).}
      \label{fig:models1}
\ifmnras
\end{figure*}
\else
\end{figure}
\fi

The pre-collapse angular velocity profile of the stellar models is worth a short discussion.
Taking the $54 M_\odot$ for example,
if the newly born NS were to rotate at approximately $\xi_\mathrm{NS}=0.1$ of its break-up speed,
the rotation speed at pre-collapse mass coordinate of $1.7 M_\odot$
and radius of $r=1395 \km$ would be approximately $\xi_0=0.008$ times its Keplerian value.
We find that if the core rotated as a solid body,
at a pre-collapse radius of
$r \approx 0.1 R_\odot$
the material is at about its break-up rotation velocity.
Furthermore, in our model the core rotates much faster than $\xi_0$,
to the degree that most of the angular momentum must be lost in the NS formation.
Therefore, a solid body rotation cannot hold for rapidly rotating
core out to $\approx R_\odot$, where the outer helium layer is
(or if there is no helium, the CO outer boundary).
Figure \ref{fig:models1} shows the differential rotation in our models.


\ifmnras
	\bibliographystyle{mn2e}

\label{lastpage}

\end{document}